\documentclass[preprint,showpacs,preprintnumbers,amsmath,amssymb]{revtex4}

\pdfoutput=1

\usepackage{graphicx}
\usepackage{dcolumn}
\usepackage{bm}
\usepackage{amssymb}

\begin{document}
\title{
Wavepacket dynamics in a quantum double-well system with \\
the Razavy potential coupled to a harmonic oscillator
}

\author{Hideo Hasegawa}
\altaffiliation{hideohasegawa@goo.jp}
\affiliation{Department of Physics, Tokyo Gakugei University,  
Koganei, Tokyo 184-8501, Japan}%

\date{\today}
\begin{abstract}
We have studied wavepacket dynamics in the Razavy hyperbolic double-well (DW) potential  
which is coupled to a harmonic oscillator (HO) by linear and quadratic interactions.
Taking into account the lowest two states of DW and $(N+1)$ states of HO ($N=1$ to 10), 
we evaluate eigenvalues and eigenfunctions of the composite system.
An analytical calculation is made for $N=1$ and numerical calculations
are performed for $1 < N \leq 10$. 
Quantum tunneling of wavepackets is realized between two bottoms 
of composite potential $U(x,y)$ 
where $x$ and $y$ denote coordinates in DW and HO potentials, respectively. 
It has been shown that with increasing $N$ and/or the coupling strength, 
the tunneling period is considerably increased.
Phase space plots of $\langle x \rangle$ vs. $\langle p_x \rangle$
and  $\langle y \rangle$ vs. $\langle p_y \rangle$ are elliptic,
where $\langle \cdot \rangle$ denotes an expectation value for the two-term wavepacket.
This result is quite different from the relevant one previously obtained 
for the quartic DW potential with the use of the quantum phase space representation
[Babyuk, arXiv:0208070].
Similarity and difference between results calculated for linear and quadratic 
couplings, and the uncertainty relation in the model are discussed.

\vspace{0.5cm}
\noindent
arXiv:1403.2368

\vspace{0.5cm}

\end{abstract}

        

\maketitle
\newpage
\section{Introduction}
A study on quantum double-well (DW) systems coupled to harmonic oscillators (HOs)
has been made in many fields of physics and chemistry \cite{Tannor07}.
Coupled DW plus HO systems have been investigated by using various methods
such as the perturbation theory \cite{Christoffel81},
time-dependent self-consistent field approximations \cite{Makri87b},
the path-integral method \cite{PI}
and the quantum phase space representation \cite{Babyuk02}.
Theoretical studies on this subject have conventionally adopted quartic potentials
for DW systems. 
However, one cannot obtain exact eigenvalues and eigenfunctions
of the Schr\"{o}dinger equation even for quartic DW potential only (without HO).
One has to apply various approximate approaches 
to quartic DW potential models.
It is furthermore difficult to obtain definite result for the coupled DW plus HO system
in which couplings between DW and HO yield an additional difficulty.

The quasi-exactly solvable hyperbolic DW potential was proposed by Razavy \cite{Razavy80}
who exactly determined a part of whole eigenvalues and eigenfunctions.
A family of quasi-exactly solvable potentials has been investigated
\cite{Finkel99,Bagchi03}.
In the present study, we adopt a DW system with the Razavy hyperbolic potential,
which is coupled to HO. 
One of advantages of our adopted model
is that we may use quasi-exactly solved eigenvalues and eigenfunctions of the DW system
with which dynamical properties of the coupled DW plus HO system may be studied.
We will consider ground and first-excited states of the DW system
which are coupled with $(N+1)$ states of HO ($N=1-10$) by linear and quadratic interactions.
In the case of $N=1$, we may make exact analytical calculations of eigenvalue and eigenfunctions
of the composite system, although we have to rely on numerical evaluation in the case of $N > 1$.
Quite recently we have studied coupled DW systems (two qubits), 
each of which is described by the Razavy potential \cite{Hasegawa14}. 
By exact calculations of eigenvalues and eigenfunctions, dynamical properties 
of coupled two DW systems have been successfully investigated \cite{Hasegawa14}.
It is worthwhile and indispensable to study wavepacket dynamics 
in quantum coupled DW plus HO system because it is a fundamental but unsettled subject.

The paper is organized as follows.
In Sec. II, we mention the calculation method employed in our study,
briefly explaining the Razavy potential \cite{Razavy80}.
Model calculations of wavepacket dynamics for linear and quadratic couplings 
with $N=1$ are presented in Secs. III A and III B, respectively.
In Sec. IV, we study motion of wavepackets including four terms, investigate
effects of adopted model parameters on the tunneling period, and present some
numerical results for the case of $1 < N \leq 10$. The uncertainty relation 
in the coupled system is also studied. Sec. V is devoted to our conclusion.

\section{The adopted method}
\subsection{Coupled double-well system with the Razavy potential}
We consider a coupled DW system whose Hamiltonian is given by 
\begin{eqnarray}
H &=& \frac{p_x^2}{2 M} +V(x) + \frac{p_y^2}{2 m} 
+\frac{m \omega^2 y^2}{2} - c\: x^{d} y,
%
\label{eq:A1}
\end{eqnarray}
with
\begin{eqnarray}
V(x) &=& \frac{\hbar^2}{2 M}
\left[\frac{\xi^2}{8} \:{\rm cosh} \:4x - 4 \xi \:{\rm cosh} \:2x- \frac{\xi^2}{8}
\right],
\label{eq:A2}
\end{eqnarray}
where $x$ ($y$) stands for coordinate of a particle of mass $M$ ($m$)
in DW (HO) potential; $p_x$ ($p_y$) means relevant momentum;
$V(x)$ signifies the Razavy DW potential \cite{Razavy80};
$\omega$ expresses the oscillator frequency of HO;
and DW and HO are coupled by linear ($d=1$) and quadratic ($d=2$)
couplings with an interaction strength of $c$.
The Razavy potential $V(x)$ with adopted parameters of $M=\xi=\hbar=1.0$ is plotted in Fig. 1(a).
Minima of $V(x)$ locate at $x_s=\pm 1.38433$ with $V(x_s)=-8.125$
and its maximum is $V(0)=-2.0$ at $x=0$.

First we consider the case of $c=0.0$ in Eq. (\ref{eq:A1}).
Eigenvalues of a DW system with the Razavy DW potential [Eq. (\ref{eq:A2})]
are given by \cite{Razavy80}
\begin{eqnarray}
\epsilon_0 &=& \frac{\hbar^2}{2M}\left[ -\xi -5 -2 \sqrt{4-2 \xi+\xi^2} \right], \\
\epsilon_1 &=& \frac{\hbar^2}{2M}\left[ \xi-5 -2 \sqrt{4+2 \xi+\xi^2} \right], \\
\epsilon_2 &=& \frac{\hbar^2}{2M}\left[ -\xi-5 +2 \sqrt{4-2 \xi+\xi^2} \right], \\
\epsilon_3 &=& \frac{\hbar^2}{2M}\left[ \xi-5 +2 \sqrt{4+2 \xi+\xi^2} \right]. 
\end{eqnarray}
Eigenvalues for the adopted parameters are $\epsilon_0=-4.73205$, $\epsilon_1=-4.64575$,
$\epsilon_2=-1.26795$ and  $\epsilon_3=0.645751$, which lead to
\begin{eqnarray}
\epsilon &=& \epsilon_1+\epsilon_0=-9.3778, 
\label{eq:A3a}\\
\delta &=& \epsilon_1-\epsilon_0=0.0863.
\label{eq:A3b}
\end{eqnarray}
Figure 1(a) shows that both $\epsilon_0$ and $\epsilon_1$ locate below $V(0)$
and that $\epsilon_2$ and $\epsilon_3$ are far above $\epsilon_1$. In this study,
we take into account the lowest two states of $\epsilon_0$ and $\epsilon_1$ 
whose eigenfunctions are given by \cite{Razavy80}
\begin{eqnarray}
\phi_0(x) &=& A_0 \; e^{-\xi \:{\rm cosh} \:2x/4} \left[3 \xi \:{\rm cosh} \:x
+(4-\xi+2 \sqrt{4-2 \xi+\xi^2})\: {\rm cosh}\: 3x \right], \\
\phi_1(x) &=&  A_1 \;e^{-\xi \:{\rm cosh} \:2x/4} \left[3 \xi \:{\rm sinh}\: x
+(4+\xi+2 \sqrt{4+2 \xi+\xi^2})\: {\rm sinh} \:3x \right], 
%
%
\end{eqnarray}
$A_\nu$ ($\nu=0, 1$) denoting normalization factors.
Figure 1(b) shows the eigenfunctions of $\phi_0(x)$ and $\phi_1(x)$, which 
are symmetric and anti-symmetric, respectively, with respect to the origin.

\begin{figure}
\begin{center}
\includegraphics[keepaspectratio=true,width=120mm]{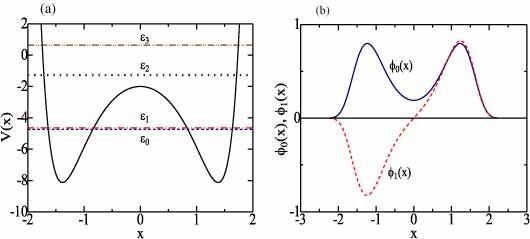}
\end{center}
\caption{
(Color online) 
(a) The Razavy DW potential $V(x)$ [Eq.(\ref{eq:A2})] 
with eigenvalues of $\epsilon_{\nu}$ ($\nu=0-3$) for $\hbar=M=\xi=1.0$.
(b) Eigenfunctions of $\phi_0(x)$ (solid curve) and $\phi_1(x)$ (dashed curve).
}
\label{fig1}
\end{figure}

The DW system in Eq. (\ref{eq:A1}) is coupled to a harmonic oscillator whose
eigenfunction and eigenvalue are given by
\begin{eqnarray}
\psi_n(y) &=& \frac{1}{\sqrt{2^n n!}} 
\left( \frac{m \omega}{\pi \hbar} \right)^{1/4}
\exp\left( -\frac{m \omega y^2}{2 \hbar}\right)
H_n\left( \sqrt{\frac{m \omega}{\hbar}}\:y \right), 
\label{eq:A4}\\
e_{n} &=& \left( n+\frac{1}{2} \right) \hbar \omega
\hspace{1cm}\mbox{($n=0,1,2\cdot,\cdot\cdot\cdot$)},
\label{eq:A5}
\end{eqnarray}
$H_n(y)$ standing for the Hermite polynomial.

\subsection{Stationary properties}
We calculate eigenvalues and eigenstates of the coupled DW system described 
by Eq. (\ref{eq:A1}). 
We expand the wavefunction with basis states of 
$\vert \nu \: n \rangle = \phi_{\nu}(x) \psi_n(y)  $ ($\nu=0, 1$ and $n=0$ to $N$) as
\begin{eqnarray}
\Phi(x, y) &=& \sum_{\nu=0}^{1} \sum_{n=0}^{N} 
\:c_{\nu, n} \:\phi_{\nu}(x) \psi_n(y),
\label{eq:A6}
\end{eqnarray}
where $N$ denotes the maximum quantum number of HO.
We obtain the secular equation
\begin{eqnarray}
E \:c_{\nu, n}=\sum_{\mu=0}^{1} \sum_{k=0}^{N} 
\:\langle \nu \:n \vert H \vert \mu\: k \rangle \;c_{\mu, k},
\label{eq:A7}
\end{eqnarray}
where 
\begin{eqnarray}
\langle \nu \:n \vert H \vert \mu\: k \rangle 
&=& \left[ \epsilon_{\nu} + \left(n+\frac{1}{2} \right) \hbar \omega \right] 
\:\delta_{\nu, \mu} \:\delta_{n, k} \nonumber \\
&-& \left[\delta_{d,1} \;\zeta \left( \delta_{\nu, \mu+1}+ \delta_{\nu, \mu-1}\right)
+ \delta_{d, 2}\;\delta_{\nu, \mu}\:\left(\zeta_0 \:\delta_{\nu,0}+\zeta_1 \:\delta_{\nu,1} \right) 
\right] \nonumber \\
& \times& \left(\sqrt{n}\: \delta_{n, k+1}+ \sqrt{n+1}\:\delta_{n, k-1} \right),
\label{eq:A8}
\end{eqnarray}
with
\begin{eqnarray}
\zeta &=& c \:\gamma \:\sqrt{\frac{g}{2}}, 
\label{eq:A9b}\\
\zeta_{\lambda} &=& c \:\gamma_{\lambda} \:\sqrt{\frac{g}{2}},
\hspace{1cm}\left( \lambda=0,1; \;\;g=\sqrt{\hbar/m \omega} \right)
\label{eq:A9}\\
\gamma &=& \int_{- \infty}^{\infty} \: \phi_0(x) \:x \:\phi_1(x)\:dx =1.13823,
\label{eq:A10}\\
\gamma_0 &=& \int_{- \infty}^{\infty}\: \phi_0(x) \:x^2 \:\phi_0(x)\:dx =1.36128, 
\label{eq:A11}\\
\gamma_1 &=& \int_{- \infty}^{\infty} \: \phi_1(x) \: x^2 \:\phi_1(x)\:dx =1.44467.
\label{eq:A12}
\end{eqnarray}
From a diagonalization of the secular equation (\ref{eq:A7}), we may obtain 
the eigenvalue $E_{\kappa}$ and eigenfunction $\Psi_{\kappa}(x, y)$
satisfying the stationary Schr\"{o}dinger equation
\begin{eqnarray}
H \Phi_{\kappa}(x,y) = E_{\kappa} \Phi_{\kappa}(x,y),
\label{eq:A12b}
\end{eqnarray}
where $\kappa=0$ to $N_m$ $=2 (N+1)-1$.
Eigenvalues and eigenfunctions for $N=1$ are analytically obtained, 
and those for $N > 1$ are evaluated by MATHEMATICA.

\subsection{Dynamical properties}
In the spectral method, a solution of the time-dependent  Schr\"{o}dinger equation
\begin{eqnarray}
i \hbar \frac{\partial \Psi(x,y,t)}{\partial t} &=& H \Psi(x,y,t),
\end{eqnarray}
is expressed by
\begin{eqnarray}
\Psi(x,y,t) &=& \sum_{\kappa=0}^{N_m}\:a_{\kappa} 
\: \Phi_{\kappa}(x,y) \:e^{-i E_{\kappa} t/\hbar},
\label{eq:A13}
\end{eqnarray}
with
\begin{eqnarray}
\sum_{\kappa=0}^{N_m} \vert a_{\kappa} \vert^2 &=& 1,
\label{eq:A13b}
\end{eqnarray}
where $E_{\kappa}$ and $\Phi_{\kappa}(x,y)$ are eigenvalue and eigenfunction,
respectively, obtained in Eq. (\ref{eq:A12b}).
Expansion coefficients $a_{\kappa}$ are in principle determined by
a given initial wavepacket, which requires cumbersome
calculations. Instead we adopt in this study, a conventional wavepacket
with coefficients given by
$a_0=a_1=1/\sqrt{2}$ and $a_{\kappa}=0$ for $\kappa \geq 2$,
\begin{eqnarray}
\Psi(x,y,t) &=& \frac{1}{\sqrt{2}}\left[ \Phi_0(x,y)\:e^{-i E_0 t/\hbar}
+ \Phi_1(x,y)\:e^{-i E_1 t/\hbar} \right].
\label{eq:A14}
\end{eqnarray}
The tunneling period $T$ for the wavepacket given by Eq. (\ref{eq:A14}) is determined by
\begin{eqnarray}
T &=& \frac{2 \pi \hbar}{E_1-E_0} =\frac{2 \pi}{\Omega_1},
\label{eq:A16}
\end{eqnarray}
where $\Omega_1=(E_1-E_0)/\hbar$. We will study 
a wavepacket with $a_0=a_1=a_2=a_3=1/2$ in Sec. IV A, whose tunneling period
is not given by Eq. (\ref{eq:A16}).

\section{Model calculations}
Introducing a parameter $\alpha$, we express the harmonic oscillator frequency $\omega$ by
\begin{eqnarray}
\hbar \omega &=& \alpha (\epsilon_1-\epsilon_0) = \alpha \:\delta.
\label{eq:B0}
\end{eqnarray}
Coefficients of $\zeta$, $\zeta_0$ and $\zeta_1$ in Eqs. (\ref{eq:A9b}) and (\ref{eq:A9}) 
are expressed in terms of $m$, $\alpha$ and $c$ as follows: 
\begin{eqnarray}
\zeta &=& c \:\gamma \left(\frac{\hbar^2}{4 m \alpha \delta}\right)^{1/4}
= c \left( \frac{\gamma \sqrt{\hbar}}{\sqrt{2} \: \delta^{1/4}} \right) 
\:\left(\frac{1}{m \alpha}\right)^{1/4}
=1.485 \:c \:\left(\frac{1}{m \alpha}\right)^{1/4}, 
\label{eq:B01}\\
\zeta_0 &=& c \left( \frac{\gamma_0 \sqrt{\hbar}}{\sqrt{2} \: \delta^{1/4}} \right) 
\:\left(\frac{1}{m \alpha}\right)^{1/4}
= 1.776 \:c \:\left(\frac{1}{m \alpha}\right)^{1/4}, 
\label{eq:B02}\\
\zeta_1 &=& c \left( \frac{\gamma_1 \sqrt{\hbar}}{\sqrt{2} \: \delta^{1/4}} \right) 
\:\left(\frac{1}{m \alpha}\right)^{1/4}
=1.885 \:c \:\left(\frac{1}{m \alpha}\right)^{1/4}.
\label{eq:B03}
\end{eqnarray}
Table 1 summarizes various coefficients appearing in our model calculations.
$\delta$, $\gamma$, $\gamma_0$, $\gamma_1$ and $\eta$ are determined by the Razavy potential
with $M=\xi=1.0$ whereas $\gamma_y$ and $\eta_y$ are given by HO potential with $m=1.0$ 
and $\omega= \alpha \delta$ ($\alpha=10.0$).
Then model calculations to be reported will be specified by a set of parameters of
$m$, $\alpha$, $c$ and $N$.

\begin{center}
\begin{tabular}[t]{|c|c|c|c|}
\hline
Coefficient & Definition & Value    &  Note  \\
\hline
$\;\;\;\delta \;\;\;$ & $\epsilon_1-\epsilon_0$ & $\;\;0.08630 \;\;$ & $\;\;\;$Eq. (\ref{eq:A3b}) $\;\;\;$   \\ 
$\;\;\;\gamma \;\;\;$ & $\langle \phi_0 \:x\: \phi_1 \rangle_x $ & $\;\;1.1382 \;\;$ & Eq. (\ref{eq:A10})    \\ 
$\gamma_0$ & $\;\; \langle \phi_0 \:x^2 \:\phi_0 \rangle_x \;\; $ & $1.3613$ & Eq. (\ref{eq:A11}) \\  
$\gamma_1$ & $\langle \phi_1 \:x^2 \:\phi_1 \rangle_x $ & $1.4447$ & Eq. (\ref{eq:A12})   \\  
$\eta$ & $\langle \phi_0 \:\partial_x\phi_1 \rangle_x $ & $0.09823$ & Eq. (\ref{eq:B13})  \\ 
\hline
$\gamma_y$ & $\langle \psi_0\:y \:\psi_1 \rangle_y $ & $0.76117$ & Eq. (\ref{eq:B14})  \\   
$\eta_y$ & $\langle \psi_0 \:\partial_y\psi_1 \rangle_y $ & $0.65689$ & Eq. (\ref{eq:B15})  \\
\hline
\end{tabular}
\end{center}
{\it Table 1} Various coefficients in model calculations
with $M=\xi=m=\hbar=1.0$ and $\alpha=10.0$,
$\langle \cdot \rangle_x$ and $\langle \cdot \rangle_y$ denoting
integrals over $x$ and $y$, respectively (see text).

\begin{figure}
\begin{center}
\includegraphics[keepaspectratio=true,width=150mm]{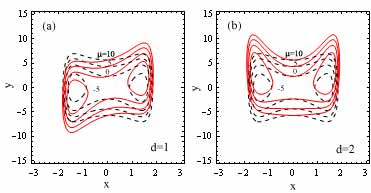}
\end{center}
\caption{
(Color online) 
Contour plots of $U(x, y)=\mu$ with $\mu=-5$, $0$, $5$ and $10$ (from the inside)
for (a) linear ($d=1$) and (b) quadratic ($d=2$) couplings 
with $c=0.0$ (dashed curves) and $c=1.0$ (solid curves).
}
\label{fig2}
\end{figure}

Figures \ref{fig2}(a) and \ref{fig2}(a) show contour maps of the composite potential
$U(x, y)=\mu$ defined by
\begin{eqnarray}
U(x, y) &=& V(x)+ \frac{m \omega^2 y^2}{2} -c \:x^{d} \;y,
\label{eq:A15} 
\end{eqnarray}
for linear ($d=1$) and quadratic ($d=2$) couplings, respectively,
with $c=0.0$ (dashed curves) and $c=1.0$ (solid curves) 
for $\mu=-5.0$, 0.0, 5.0 and 10.0 ($m=1.0$ and $\alpha=10.0$).
For $c=0$, $U(x, y)$ has two minima of $U(x, y)=-8.125$ at $(x, y)=(\pm 1.3843, \;0.0)$.
For a linear coupling ($d=1$) with $c=1.0$, 
it has two minima of $U(x, y)= -9.438$ at $(x, y)=(1.4120,\; 1.8959)$
and $(-1.4120,\; -1.8959)$.
For a quadratic coupling ($d=2$) with $c=1.0$,
it has two minima of $U(x, y)= -9.125$ at $(x, y)=(1.4120,\; 1.8959)$
and $(-1.4120,\; 1.8959)$.
Model calculations for linear and quadratic couplings with $N=1$ will be separately
reported in Secs. III A and III B, respectively.

\subsection{Linear coupling with $N=1$}
\begin{figure}
\begin{center}
\includegraphics[keepaspectratio=true,width=120mm]{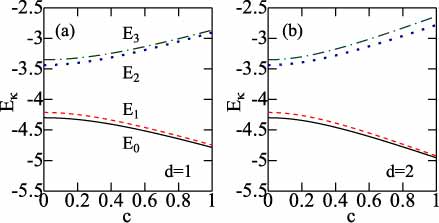}
\end{center}
\caption{
(Color online) 
The $c$ dependence of eigenvalues $E_{\kappa}$ ($\kappa=0-$3) 
of (a) a linear coupling ($d=1$) and (b) a quadratic coupling ($d=2$)
with $m=1.0$, $\alpha=10.0$ and $N=1$.
}
\label{fig3}
\end{figure}

\begin{figure}
\begin{center}
\includegraphics[keepaspectratio=true,width=120mm]{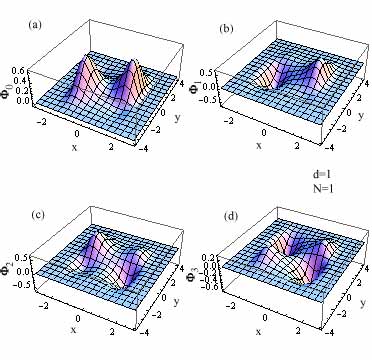}
\end{center}
\caption{
(Color online) 
Eigenfunctions of (a) $\Phi_0(x,y)$, (b) $\Phi_1(x,y)$, (c) $\Phi_2(x,y)$,
and (d) $\Phi_3(x,y)$ for a linear coupling ($d=1$) 
($m=1.0$, $\alpha=10.0$, $c=1.0$ and $N=1$).
}
\label{fig4}
\end{figure}

For a linear coupling ($d=1$) with $N=1$, the energy matrix of the Hamiltonian 
given by Eq. (\ref{eq:A1}) is expressed in the basis of $\psi_0(y) \phi_0(x)$,  $\psi_0(y) \phi_1(x) $, 
$ \psi_1(y) \phi_0(x)$ and $\psi_1(y) \phi_1(x)$ by
\begin{eqnarray}
{\cal H} &=& \left( {\begin{array}{*{20}c}
   {\epsilon_0 +\hbar \omega/2} & {0 } & {0 } & {- \zeta} \\
   {0 } & {\epsilon_1 + \hbar \omega/2 } & {-\zeta } & {0} \\
   {0 } & {- \zeta } & {\epsilon_0 + 3 \hbar \omega/2 } & {0} \\
   {- \zeta } & {0 } & {0 } & {\epsilon_1+ 3 \hbar \omega/2} \\   
\end{array}} \right),
\label{eq:B1}
\end{eqnarray}
where $\zeta$ is given by Eq. (\ref{eq:B01}).
We obtain eigenvalues of the energy matrix
\begin{eqnarray}
E_0 &=& \frac{\epsilon}{2}+ \hbar \omega -\sqrt{\frac{1}{4}(\hbar \omega+\delta)^2+ \zeta^2}, 
\label{eq:B2}\\
E_1 &=& \frac{\epsilon}{2}+ \hbar \omega -\sqrt{\frac{1}{4}(\hbar \omega-\delta)^2+ \zeta^2}, \\
E_2 &=& \frac{\epsilon}{2}+ \hbar \omega +\sqrt{\frac{1}{4}(\hbar \omega-\delta)^2+ \zeta^2}, \\
E_3 &=& \frac{\epsilon}{2}+ \hbar \omega +\sqrt{\frac{1}{4}(\hbar \omega+\delta)^2+ \zeta^2}.
\label{eq:B3}
\end{eqnarray}
Relevant eigenfunctions are expressed by
\begin{eqnarray}
\Phi_0(x, y) &=& \cos \theta_1 \: \psi_0(y) \phi_0(x) 
+ \sin \theta_1 \: \psi_1(y)\phi_1(x), 
\label{eq:B4}\\
\Phi_1(x, y) &=& \cos \theta_2 \:  \psi_0(y) \phi_1(x)
+ \sin \theta_2 \: \psi_1(y) \phi_0(x), \\
\Phi_2(x, y) &=& -\sin \theta_2 \: \psi_0(y) \phi_1(x)
+ \cos \theta_2 \: \psi_1(y) \phi_0(x), \\
\Phi_3(x, y) &=& -\sin \theta_1 \:\psi_0(y) \phi_0(x) 
+ \cos \theta_1 \: \psi_1(y) \phi_1(x),
\label{eq:B5}
\end{eqnarray}
where
\begin{eqnarray}
\tan \:2 \theta_1 &=& \frac{2 \zeta}{(\hbar \omega+\delta)}, 
\label{eq:B6a}\\
\tan \:2 \theta_2 &=& \frac{2 \zeta}{(\hbar \omega-\delta)}.
\label{eq:B6}
\end{eqnarray}
Eigenvalues $E_{\kappa}$ ($\kappa=0-3$) for $d=1$ with $m=1.0$ and $\alpha=10.0$
are plotted as a function of $c$ in Fig. \ref{fig3}(a):
Fig. \ref{fig3}(b) for $d=2$ will be explained later (Sec. III B).
An energy gap between the ground and first-excited states is
$\Omega_1=0.08630$ for $c=0.0$ and $\Omega_1=0.03958$ for $c=1.0$.
$\Omega_1$ is decreased with increasing $c$.
Figures \ref{fig4}(a)-\ref{fig4}(d) show 3D plots of eigenfunctions of
$\Phi_{\kappa}(x,y)$ ($\kappa=0-3$). 

\begin{figure}
\begin{center}
\includegraphics[keepaspectratio=true,width=150mm]{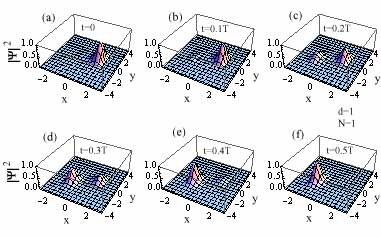}
\end{center}
\caption{
(Color online) 
$\vert \Psi(x,y,t) \vert^2$ for a linear coupling ($d=1$) 
at (a) $t=0.0$, (b) $t=0.1 T$, (c) $t=0.2 T$, (d) $t=0.3 T$, 
(e) $t=0.4 T$ and (f) $t= 0.5 T$ where $T=158.73$ ($m=1.0$, $\alpha=10.0$, $c=1.0$ and $N=1$).
}
\label{fig5}
\end{figure}

We investigate motion of a wavepacket consisting of $\Phi_0(x,y)$ and $\Phi_1(x,y)$
given by Eq. (\ref{eq:A14}).
Time-dependent wavepackets are illustrated in
Figs. \ref{fig5}(a)-(f) which show 3D plots of $\vert \Psi(x,y,t) \vert^2$
at (a) $t=0.0$, (b) $0.1 T$, (c) $0.2 T$, (d) $0.3 T$, (e) $0.4 T$ and (f) $0.5 T$,
where $T=158.73$ obtained by $\Omega_1=0.03958$.
Wavepackets at $t=0.6 T$, $0.7 T$, $0.8 T$, $0.9 T$ and $T$ are the same as
those at $t=0.4 T$, $0.3 T$, $0.2 T$, $0.1 T$ and $0.0$, respectively.
At $t=0.0$, a peak of the wavepacket locates at $(x_m, y_m)=(1.2353, 0.61990)$.
With time developing, a peak of the wavepacket at $(x, y)=(-1.2353, -0.61990)$ is growing,
and it goes back to the initial position at $t=T$.
The wavepacket shows a tunneling from $(x, y)=(1.2353, 0.61990)$
to $(x, y)=(-1.2353, -0.61990)$ across the potential barrier at the origin 
[see Fig. \ref{fig2}(a)].

\begin{figure}
\begin{center}
\includegraphics[keepaspectratio=true,width=150mm]{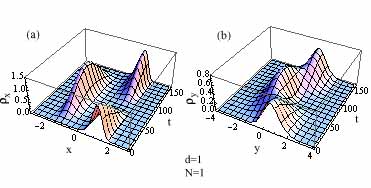}
\end{center}
\caption{
(Color online) 
Time dependence of (a) $\rho_x(t)$ and (b) $\rho_y(t)$ for a linear coupling ($d=1$)
($m=1.0$, $\alpha=10.0$, $c=1.0$ and $N=1$).
}
\label{fig6}
\end{figure}

\begin{figure}
\begin{center}
\includegraphics[keepaspectratio=true,width=110mm]{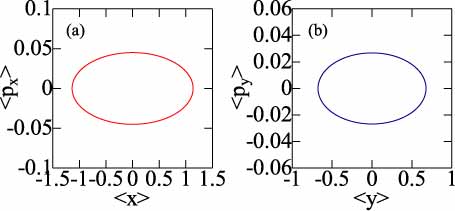}
\end{center}
\caption{
(Color online) 
Phase space representations of (a) $ \langle p_x \rangle $ vs.  $ \langle x \rangle $ 
and (b) $ \langle p_y \rangle $ vs. $ \langle y \rangle $ for a linear coupling ($d=1$)
($m=1.0$, $\alpha=10.0$, $c=1.0$ and $N=1$).
}
\label{fig7}
\end{figure}

By using Eqs. (\ref{eq:A14}), (\ref{eq:B4})-(\ref{eq:B5}),
we may calculate marginal probability densities of $x$ and $y$ components,
which are given by
\begin{eqnarray}
\rho_x(t) &=& \int_{-\infty}^{\infty}\:\vert \Psi(x,y,t) \vert^2 \:dy, \\
&=& \frac{1}{2}\left( \cos^2 \theta_1+ \sin^2 \theta_2 \right) \phi_0(x)^2
+ \frac{1}{2}\left( \sin^2 \theta_1+ \cos^2 \theta_2 \right) \phi_1(x)^2 \nonumber \\
&+& \cos (\theta_1-\theta_2) \phi_0(x) \phi_1(x) \: \cos \Omega_1 t, 
\label{eq:B7}\\
\rho_y(t) &=& \int_{-\infty}^{\infty} \:\vert \Psi(x,y,t) \vert^2 \:dx, \\
&=& \frac{1}{2}\left( \cos^2 \theta_1+ \cos^2 \theta_2 \right) \psi_0(y)^2
+ \frac{1}{2}\left( \sin^2 \theta_1+ \sin^2 \theta_2 \right) \psi_1(y)^2 \nonumber \\
&+& \sin(\theta_1+\theta_2) \psi_0(y) \psi_1(y) \: \cos \Omega_1 t.
\label{eq:B8}
\end{eqnarray}
Figures \ref{fig6}(a) and \ref{fig6}(b) show $\rho_x(t)$ and $\rho_y(t)$, respectively.
Both $\rho_x(t)$ and $\rho_y(t)$ oscillate with the same period.

The tunneling probability of $P_r(t)$ for finding a particle in the negative $x$ region is given by
\begin{eqnarray}
P_r(t) &=& \int_{-\infty}^0 \: \rho_x(t) \:dx, \\
&=&\frac{1}{2} - b\: \cos (\theta_1-\theta_2) \: \cos \Omega_1 t, 
\label{eq:B8b}
\end{eqnarray}
with
\begin{eqnarray}
b &=& - \int_{-\infty}^0 \: \phi_0(x) \phi_1(x) \:dx =0.496213.
\end{eqnarray}

By simple calculations, we obtain various time-dependent expectation values 
given by
\begin{eqnarray}
\langle x \rangle 
&=& \int_{-\infty}^{\infty} \int_{-\infty}^{\infty} 
\: \Psi^*(x,y,t) \: x \:  \Psi(x,y,t)  \:dx \:dy, \\
&=& \gamma \:\cos(\theta_1-\theta_2)\:\cos \Omega_1 t, 
\label{eq:B9}\\
\langle p_x \rangle  
&=& \int_{-\infty}^{\infty} \int_{-\infty}^{\infty} 
\: \Psi^*(x,y,t) \: (-i \partial_x)\: \Psi(x,y,t)  \:dx \: dy, \\
&=& -\eta \:\cos(\theta_1+\theta_2)\:\sin \Omega_1 t, 
\label{eq:B10}\\
\langle y \rangle &=& \gamma_y \:\sin(\theta_1+\theta_2)\:\cos \Omega_1 t, 
\label{eq:B11}\\
\langle p_y \rangle &=& \eta_y \:\sin(\theta_1-\theta_2)\:\sin \Omega_1 t,
\label{eq:B12}
\end{eqnarray}
with
\begin{eqnarray}
\eta &=& \int_{-\infty}^{\infty} \: \phi_0(x) \:\partial_x \phi_1(x)\:dx
=- \int_{-\infty}^{\infty} \:\phi_1(x) \:\partial_x \phi_0(x)\:dx=0.09823, 
\label{eq:B13}\\
\gamma_y &=&  \int_{-\infty}^{\infty} \: \psi_0(y) \:y \: \psi_1(y)\:dy=0.76117, 
\label{eq:B14}\\
\eta_y &=& \int_{-\infty}^{\infty} \: \psi_0(y) \:\partial_y \psi_1(y)\:dy
=- \int_{-\infty}^{\infty} \: \psi_1(y) \:\partial_y\psi_0(y)\:dy=0.65689,
\label{eq:B15}
\end{eqnarray}
where $\gamma$ is given by Eq. (\ref{eq:A10}).
We generally observe that 
$\langle p_x \rangle=M \langle dx/dt \rangle \neq M d\langle x \rangle/d t$
because of the nonlinearlity of the adopted system.
Parametric plots of both $\langle x \rangle$ vs. $\langle p_x \rangle$ 
and $\langle y \rangle$ vs. $\langle p_y \rangle$
are elliptic, as shown in Figs. \ref{fig7}(a) and \ref{fig7}(b).

\subsection{Quadratic coupling with $N=1$}
\begin{figure}
\begin{center}
\includegraphics[keepaspectratio=true,width=120mm]{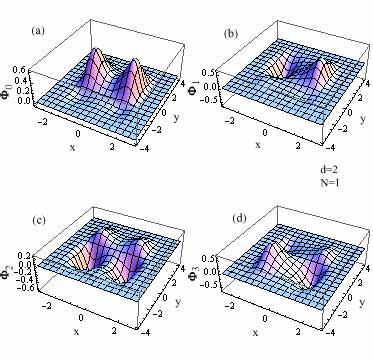}
\end{center}
\caption{
(Color online) 
Eigenfunctions of (a) $\Phi_0(x,y)$, (b) $\Phi_1(x,y)$, (c) $\Phi_2(x,y)$,
and (d) $\Phi_3(x,y)$ for the quadratic coupling ($d=2$) for a quadratic
coupling ($d=2$) ($m=1.0$, $\alpha=10.0$, $c=1.0$ and $N=1$).
}
\label{fig8}
\end{figure}

Next we consider a quadratic coupling ($d=2$), for which the energy matrix with $N=1$ 
is expressed in the basis of $\psi_0(y)\phi_0(x)$, $\psi_0(y)\phi_1(x)$, $\psi_1(y)\phi_0(x)$ 
and $\psi_1(y)\phi_1(x)$ by
\begin{eqnarray}
{\cal H} &=& \left( {\begin{array}{*{20}c}
   {\epsilon_0 +\hbar \omega/2} & {0 } & {- \zeta_0 } & {0} \\
   {0 } & {\epsilon_1 + \hbar \omega/2 } & {0} & {-\zeta_1 } \\
   {- \zeta_0 } & {0 } & {\epsilon_0 + 3 \hbar \omega/2 } & {0} \\
   {0} & {- \zeta_1 } & {0 } & {\epsilon_1+ 3 \hbar \omega/2} \\   
\end{array}} \right),
\label{eq:C1}
\end{eqnarray}
where $\zeta_0$ and $\zeta_1$ are given by Eqs. (\ref{eq:B02}) and (\ref{eq:B03}),
respectively. We obtain eigenvalues of the energy matrix given by
\begin{eqnarray}
E_0 &=& \epsilon_0+ \hbar \omega -\sqrt{\frac{1}{4}(\hbar \omega)^2+ \zeta_0^2}, 
\label{eq:C2}
\\
E_1 &=& \epsilon_1+ \hbar \omega -\sqrt{\frac{1}{4}(\hbar \omega)^2+ \zeta_1^2}, \\
E_2 &=& \epsilon_0+ \hbar \omega +\sqrt{\frac{1}{4}(\hbar \omega)^2+ \zeta_0^2}, \\
E_3 &=& \epsilon_1+ \hbar \omega +\sqrt{\frac{1}{4}(\hbar \omega)^2+ \zeta_1^2}.
\label{eq:C3}
\end{eqnarray}
Relevant eigenfunctions are expressed by
\begin{eqnarray}
\Phi_0(x, y) &=& \cos \theta_1 \: \psi_0(y) \phi_0(x) 
+ \sin \theta_1 \: \psi_1(y)\phi_0(x), 
\label{eq:C4}
\\
\Phi_1(x, y) &=& \cos \theta_2 \:  \psi_0(y) \phi_1(x)
+ \sin \theta_2 \: \psi_1(y) \phi_1(x), \\
\Phi_2(x, y) &=& -\sin \theta_1 \: \psi_0(y) \phi_0(x)
+ \cos \theta_1 \: \psi_1(y) \phi_0(x), \\
\Phi_3(x, y) &=& -\sin \theta_2 \:\psi_0(y) \phi_1(x) 
+ \cos \theta_2 \: \psi_1(y) \phi_1(x),
\label{eq:C5}
\end{eqnarray}
where
\begin{eqnarray}
\tan \:2 \theta_1 &=& \frac{2 \zeta_0}{\hbar \omega}, 
\label{eq:C6a} \\
\tan \:2 \theta_2 &=& \frac{2 \zeta_1}{\hbar \omega}.
\label{eq:C6}
\end{eqnarray}
Eigenvalues $E_{\kappa}$ ($\kappa=0-3$) for $d=2$ with $m=1.0$ and $\alpha=10.0$ 
are plotted as a function of $c$ in Fig. \ref{fig3}(b).
An energy gap between the ground and first-excited states is
$\Omega_1=0.08630$ and $0.02989$ for $c=0.0$ and $1.0$, respectively. 
The $c$ dependence of eigenvalues for a quadratic coupling
is similar to that for a linear coupling shown in Fig. \ref{fig3}(a).
Figures \ref{fig8}(a)-\ref{fig8}(d) show eigenfunctions 
of $\Phi_{\kappa}(x,y)$ ($\kappa=0-3$).

\begin{figure}
\begin{center}
\includegraphics[keepaspectratio=true,width=150mm]{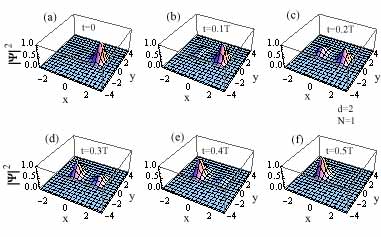}
\end{center}
\caption{
(Color online) 
$\vert \Psi(x,y,t) \vert^2$ for a quadratic coupling ($d=2$)
at (a) $t=0.0$, (b) $t=0.1 T$, (c) $t=0.2 T$, (d) $t=0.3 T$, (e) $t=0.4 T$ and (f) $t=0.5 T$
where $T=210.25$ for a quadratic coupling ($d=2$) ($m=1.0$, $\alpha=10.0$, $c=1.0$ and $N=1$).
}
\label{fig9}
\end{figure}

We consider a wavepacket $\Psi(x,y,t)$ 
given by Eq. (\ref{eq:A14}).
Figures \ref{fig9}(a)-\ref{fig9}(f) show 3D plots of $\vert \Psi(x,y,t) \vert^2$
at (a) $T=0$, (b) $0.1 T$, (c) $0.2 T$, (d) $0.3 T$, (e) $0.4 T$ and (f) $0.5 T$
where $T=210.25$.
At $t=0.0$, a peak of the wavepacket locates at $(x_m, y_m)=(1.2354, 0.6468)$.
At $t \sim 0.5 \;T$, wavepacket has appreciable magnitude at $(x, y)=(-1.2354, 0.6468)$
because the minimum of the composite potential $U(x,y)$ locates at $(x, y)=(-1.4120,\; 1.8959)$.
The tunneling of the wavepacket occurs between 
$(x, y)=(1.2354, 0.6468)$ and $(-1.2354, 0.6468)$.

\begin{figure}
\begin{center}
\includegraphics[keepaspectratio=true,width=150mm]{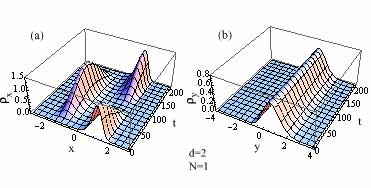}
\end{center}
\caption{
(Color online) 
Time dependence of (a) $\rho_x(t)$ and (b) $\rho_y(t)$ for a quadratic coupling ($d=2$)
($m=1.0$, $\alpha=10.0$, $c=1.0$ and $N=1$).
}
\label{fig10}
\end{figure}

\begin{figure}
\begin{center}
\includegraphics[keepaspectratio=true,width=110mm]{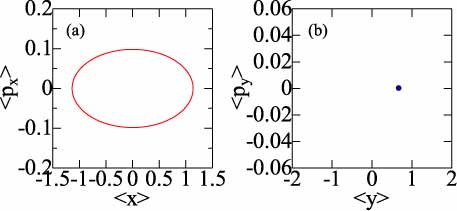}
\end{center}
\caption{
(Color online) 
Phase space representations of (a) $ \langle p_x \rangle $ vs.  $ \langle x \rangle $ 
and (b) $ \langle p_y \rangle $ vs. $ \langle y \rangle $ for a quadratic coupling ($d=2$)
($m=1.0$, $\alpha=10.0$, $c=1.0$ and $N=1$).
}
\label{fig11}
\end{figure}

By using Eqs. (\ref{eq:A14}), (\ref{eq:B4})-(\ref{eq:B5}), we may obtain
marginal probability densities of $x$ and $y$ components, which are given by
\begin{eqnarray}
\rho_x(t) &=& \frac{1}{2}\left[\phi_0(x)^2+ \phi_1(x)^2 \right]
+ \cos (\theta_1-\theta_2) \:\phi_0(x) \phi_1(x) \: \cos \Omega_1 t,
\label{eq:C7}\\
\rho_y(t) &=&  \frac{1}{2} \left[ \left( \cos^2 \theta_1+ \cos^2 \theta_2 \right) \psi_0(y)^2
+ \left( \sin^2 \theta_1+ \sin^2 \theta_2 \right) \psi_1(y)^2 \right] \nonumber \\
&+& \frac{1}{2}\left[ \sin(2 \theta_1)+ \sin(2 \theta_2) \psi_0(y) \psi_1(y) \right].
\label{eq:C8}
\end{eqnarray}
Figures \ref{fig10}(a) and \ref{fig10}(b) show $\rho_x(t)$ and $\rho_y(t)$, respectively.
$\rho_x(t)$ is similar to the relevant result for the linear coupling in Fig. \ref{fig6}(a) 
although $\rho_y(t)$ is different from that in Fig. \ref{fig6}(b).

The tunneling probability $P_r(t)$ is given by
\begin{eqnarray}
P_r(t) &=& \frac{1}{2}-b \cos{(\theta_1-\theta_2)} \cos \Omega_1 t,
\end{eqnarray}
which is the same as Eq. (\ref{eq:B8b}) for a linear coupling.

Various time-dependent expectation values 
are given by
\begin{eqnarray}
\langle x \rangle 
&=& \gamma \:\cos(\theta_1-\theta_2)\:\cos \Omega_1 t, 
\label{eq:C9}\\
\langle p_x \rangle  
&=& -\eta \:\cos(\theta_1-\theta_2)\:\sin \Omega_1 t, 
\label{eq:C10}\\
\langle y \rangle &=& \frac{\gamma_y}{2} \left( \:\sin 2\theta_1+ \sin 2\theta_2 \right), 
\label{eq:C11}\\
\langle p_y \rangle &=& 0.
\label{eq:C12}
\end{eqnarray}
Figures \ref{fig11}(a) and \ref{fig11}(b) show parametric plots of
$\langle x \rangle$ vs. $\langle p_x \rangle$ and $\langle y \rangle$ vs. $\langle p_y \rangle$,
respectively. The former is ellipsoid while the latter is a point at
$(\langle y \rangle, \langle p_y \rangle)=(0.70186, 0.0)$ staying at the initial state.
Although $\langle x \rangle$ vs. $\langle p_x \rangle$ plot in Fig. \ref{fig11}(a) 
is similar to that for a linear coupling in Fig. \ref{fig7}(a),  
$\langle y \rangle$ vs. $\langle p_y \rangle$ plot in Fig. \ref{fig11}(b)
is quite different from that for a linear coupling in Fig. \ref{fig7}(b).

\begin{figure}
\begin{center}
\includegraphics[keepaspectratio=true,width=150mm]{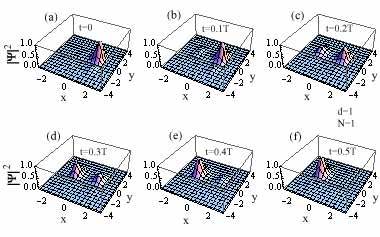}
\end{center}
\caption{
(Color online) 
$\vert \Psi(x,y,t) \vert^2$ of the four-component wavepacket given by Eq. (\ref{eq:D1})
for a linear coupling ($d=1$) at (a) $t=0.0$, (b) $t=0.1 T$, (c) $t=0.2 T$, (d) $t=0.3 T$, 
(e) $t=0.4 T$ and (f) $t= 0.5 T$ where $T=71.6084$ ($m=1.0$, $\alpha=10.0$, $c=0.1$ and $N=1$).
}
\label{fig12}
\end{figure}

\section{Discussion}
\subsection{A wavepacket with $a_0=a_1=a_2=a_3=1/2$}
In the preceding section, we consider a wavepacket with $a_0=a_1=1/\sqrt{2}$ and $a_2=a_3=0.0$.
Here we will study a four-component wavepacket with coefficients of $a_0=a_1=a_2=a_3=1/2$ 
in Eq. (\ref{eq:A13})
\begin{eqnarray}
\Psi(x,y,t) &=& \frac{1}{2}\sum_{\kappa=0}^{3}\;\Phi_{\kappa}(x,y) \:e^{-i E_{\kappa} t/\hbar}.
\label{eq:D1}
\end{eqnarray}
For a linear coupling ($d=1$) with $c=0.1$, $m=1.0$, $\alpha=10.0$ and $N=1$, 
we obtain eigenvalues of $(E_0, E_1, E_2, E_3)=(-4.30784, -4.22313, -3.42868, -3.34397)$.
The peak of the wavepacket initially locates at $(x,y)=(x_m,y_m)=(1.2353, \:0.80775)$.
The time-dependence of $\vert \Psi(x,y,t) \vert^2$ 
from $t=0$ to $t=T/2$ are shown in Fig. \ref{fig12} where $T=71.6084$ (below).
With time developing, a new peak appears at $(x,y) \neq (x_m,y_m)$, and 
at $t=T$ wavepacket returns to its initial position.

\begin{figure}
\begin{center}
\includegraphics[keepaspectratio=true,width=140mm]{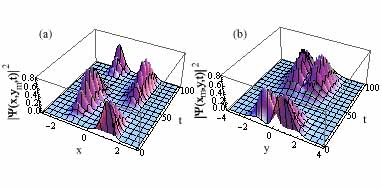}
\end{center}
\caption{
(Color online) 
(a) 3D plots of $\vert \Psi(x,y_m,t) \vert^2$ with $y_m=0.761169$ 
and (b) $\vert \Psi(x_m,y,t) \vert^2$ with $x_m=1.23534$ of the wavepacket given 
by Eq. (\ref{eq:D1}) ($m=1.0$, $\alpha=10.0$, $c=0.1$, $d=1$ and $N=1$).
}
\label{fig13}
\end{figure}

\begin{figure}
\begin{center}
\includegraphics[keepaspectratio=true,width=80mm]{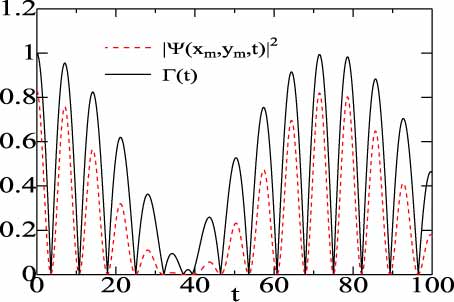}
\end{center}
\caption{
(Color online) 
Time dependence of $\Gamma(t)$ (solid curve) 
and $\vert \Psi(x_m,y_m,t) \vert^2$ (dashed curve) with $(x_m, y_m)=(1.23534, \:0.761169)$
for the wavepacket given by Eq. (\ref{eq:D1})
($m=1.0$, $\alpha=10.0$, $c=0.1$, $d=1$ and $N=1$).
}
\label{fig14}
\end{figure}

Calculations of $\rho_x(t)$ and $\rho_y(t)$ for this wavepacket 
consisting of four terms are very tedious though it is not impossible.
As their substitutes, we show the 3D plot
of $\vert \Psi(x,y_m,t) \vert^2$ as functions of $x$ and $t$ in Fig. \ref{fig13}(a),
and that of $\vert \Psi(x_m,y,t) \vert^2$ as functions of $y$ and $t$ in Fig. \ref{fig13}(b).
Both $\vert \Psi(x,y_m,t) \vert^2$ and $\vert \Psi(x_m,y,t) \vert^2$
show complicated and rapid oscillations.
The dashed curve in Fig. \ref{fig14} expresses $\vert \Psi(x_m,y_m,t) \vert^2$
as a function of $t$, and the solid curve 
shows the correlation function $\Gamma(t)$ defined by
\begin{eqnarray}
\Gamma(t) &=& \vert \int_{-\infty}^{\infty} \int_{-\infty}^{\infty} 
\:\Psi^*(x,y, 0) \:\Psi(x,y,t)\;dx dy \:\vert, \\
&=& \frac{1}{4} \vert1+ \: e^{- i \Omega_1 t}+e^{- i \Omega_2 t}+e^{- i \Omega_3 t} \vert.
\hspace{1cm}\mbox{ ($\Omega_{\kappa}=(E_{\kappa}-E_0)/\hbar) $}
\label{eq:G2}
\end{eqnarray}
From the condition for the tunneling period $T$,
\begin{eqnarray}
T = \min_{\forall \:t \:> 0}\; \large\{ \Gamma(t)=1 \large\},
\end{eqnarray}
we obtain $T=71.6084$ which is slightly different from a value estimated by $2 \pi/(E_1-E_0)=74.1706$.
Although the tunneling period is mainly determined by $E_0$ and $E_1$, 
its precise value is influenced by contributions from higher exited states with
$E_2$ and $E_3$.

\subsection{Coupling dependence of $T$ for other choices of parameters of $m$ and $\alpha$}

\begin{figure}
\begin{center}
\includegraphics[keepaspectratio=true,width=80mm]{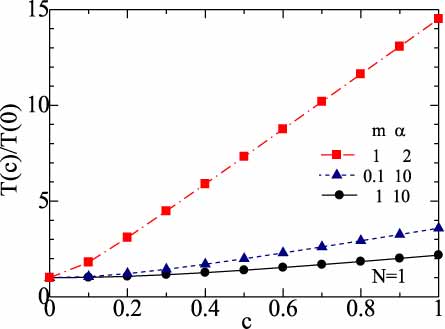}
\end{center}
\caption{
(Color online) 
The $c$ dependence of the tunneling period $T$ for three sets of parameters of
$(m, \alpha)=(1.0, 10.0)$ (solid curve), $(0.1, 10.0)$ (dashed curve) 
and $(1.0, 2.0)$ (chain curve) with $N=1$. 
}
\label{fig15}
\end{figure}

We have so far presented model calculations with a set of parameters
of $(m, \alpha)=(1.0, 10.0)$ with $c=1.0$ and $N=1$. 
We have calculated the tunneling period $T$ as a function of the interaction $c$
for three sets of parameters: $(m, \alpha)=(1.0, 10.0)$, $(0.1, 10.0)$ and $(1.0, 2.0)$,
whose results are shown in Fig. \ref{fig15}.
We note that $T$ is increased with increasing $c$, which is more significant 
for smaller $m$ and for smaller $\alpha$.

\subsection{The case of $N > 1$}
In the case of $N > 1$, we have to numerically evaluate eigenvalues and eigenfunctions
of the energy matrix with dimension of $(N_m+1) \times (N_m+1)$ by using MATHEMATICA,
where $N_m+1 =2 (N+1)$. 
Since we have neglected excited states higher than $\epsilon_2$ of DW system,
a reasonable choice of $N$ for the maximum quantum state of HO is expected to be given by
\begin{eqnarray}
\left( N+\frac{1}{2} \right) \:\hbar \omega
\sim (\epsilon_2-\epsilon_0) =3.4641.
\label{eq:C13}
\end{eqnarray}
By using Eqs. (\ref{eq:B0}) and (\ref{eq:C13}), we obtain $N \sim 3.5$ for $\alpha=10.0$ and 
$N \sim 19.6$ for $\alpha=2.0$.

\begin{figure}
\begin{center}
\includegraphics[keepaspectratio=true,width=80mm]{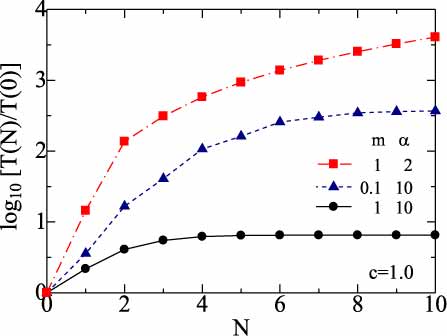}
\end{center}
\caption{
(Color online) 
The $N$ dependence of the tunneling period $T$ for three sets of parameters of
$(m, \alpha)=(1.0, 10.0)$ (solid curve), $(0.1, 10.0)$ (dashed curve) 
and $(1.0, 2.0)$ (chain curve) for a linear coupling ($d=1$) with $c=1.0$.
Results of $N=0$ stand for those of no couplings.
}
\label{fig16}
\end{figure}

$N$ dependences of the tunneling period $T$ calculated for 
three sets of parameters: $(m, \alpha)=(1.0, 10.0)$, $(0.1, 10.0)$ and $(1.0, 2.0)$ 
with $c=1.0$ are shown in Fig. \ref{fig16} where the ordinate is in the logarithmic scale.
It is noted that $T$ is significantly increased with increasing $N$, in particular
for smaller $m$ and smaller $\alpha$.
In the case of $(m, \alpha)=(1.0, 10.0)$, the enhancement of $T$ saturates
at $N \gtrsim 4$. On the contrary, such a saturation is not realized 
in the case of $(m, \alpha)=(1.0, 2.0)$ even at $N=10$.

\begin{figure}
\begin{center}
\includegraphics[keepaspectratio=true,width=150mm]{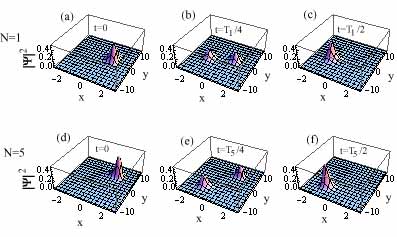}
\end{center}
\caption{
(Color online) 
$\vert \Psi(x,y,t) \vert^2$ with $N=1$  
at (a) $t=0.0$, (b) $t=T_1/4$ and (c) $t=T_1/2$ where $T_1=1056.6$;
$\vert \Psi(x,y,t) \vert^2$ with $N=5$ at (d) $t=0.0$, (e) $t=T_5/4$ and (f) $t=T_5/2$
where $T_5=68046.6$ ($d=1$, $m=1.0$, $\alpha=2.0$ and $c=1.0$).
}
\label{fig17}
\end{figure}

Paying attention to the case of $(m, \alpha)=(1.0, 2.0)$ which shows the 
most significant $N$ dependence of $T$ in Fig. \ref{fig16},
we have calculated the time-dependent wavepackets for $N=1$ and $5$,
whose results are shown in Fig. \ref{fig17}.
We obtain $\Omega_1=0.594662 \times 10^{-3}$ for $N=1$,
and $\Omega_1=0.923365 \times 10^{-4}$ for $N=5$.
The initial position of wavepacket for $N=1$ is $(x, y)=(1.23526, 1.66213)$,
while that for $N=5$ is $(x, y)=(1.23532, 5.54669)$.
Figure \ref{fig17}(a)-\ref{fig17}(c) show magnitudes of wavepackets 
at $0.0 \leq t \leq T_1/2$ where $T_1\;(=1056.6)$ denotes the tunneling period for $N=1$.
Figure \ref{fig17}(d)-\ref{fig17}(f) show similar results
at $0.0 \leq t \leq T_5/2$ where the tunneling period for $N=5$ is $T_5=68046.6$.
Comparing time-dependent magnitudes of wavepackets in Figs. \ref{fig17}(a)-\ref{fig17}(c) for $N=1$ 
with those in Fig. \ref{fig17}(d)-\ref{fig17}(f) for $N=5$, we note that two results 
are similar when reading them by the normalized time $t/T$, 
despite the fact that the tunneling period $T_5$ is larger than $T_1$ by a factor of 64.4.

\subsection{Uncertainty relation}
The Heisenberg uncertainty of $\Delta x \Delta p_x$, which is also a typical quantum phenomenon,
is related with the tunneling \cite{Hasegawa13}.
We may obtain analytical expressions for averages of fluctuations of $x$ and $p_x$
in the case of $N=1$.
For a linear coupling, we obtain

\begin{eqnarray}
(\Delta x)^2 &=& \langle x^2 \rangle- \langle x \rangle^2, \\
&=& \frac{\gamma_0}{2}(\cos^2 \theta_1+\sin^2 \theta_2)+
\frac{\gamma_1}{2}(\sin^2 \theta_1+\cos^2 \theta_2)-
\gamma^2 \cos^2 (\theta_1-\theta_2)\;\cos^2 \Omega_1 t, 
\label{eq:Z1}\\
(\Delta p_x)^2 &=& \langle p_x^2 \rangle - \langle p_x \rangle^2, \\
&=& \frac{\chi_0}{2}(\cos^2 \theta_1+\sin^2 \theta_2)+
\frac{\chi_1}{2}(\sin^2 \theta_1+\cos^2 \theta_2)-
\eta^2 \cos^2 (\theta_1+\theta_2)\;\sin^2 \Omega_1 t, 
\label{eq:Z2}
\end{eqnarray}
with
\begin{eqnarray}
\chi_0 &=& \int_{-\infty}^{\infty}\:[\partial_x \phi_0(x)]^2 \:dx =2.58707,\\
\chi_1 &=& \int_{-\infty}^{\infty} \:[\partial_x \phi_1(x)]^2 \:dx =3.17399,
\end{eqnarray}
where $\theta_1$ and $\theta_2$ are given by Eqs. (\ref{eq:B6a}) 
and (\ref{eq:B6}), respectively.
For a quadratic coupling, they are given by
\begin{eqnarray}
(\Delta x)^2 &=& \frac{\gamma_0}{2}+\frac{\gamma_1}{2}-
\gamma^2 \cos^2 (\theta_1-\theta_2)\;\cos^2 \Omega_1 t, \\
(\Delta p_x)^2 &=& \frac{\chi_0}{2}+\frac{\chi_1}{2}-
\eta^2 \cos^2 (\theta_1-\theta_2)\;\sin^2 \Omega_1 t,
\end{eqnarray}
where $\theta_1$ and $\theta_2$ are given by Eqs. (\ref{eq:C6a}) 
and (\ref{eq:C6}), respectively.
For uncoupled DW ($c=0.0$) where $\theta_1=\theta_2=0$, Eqs. (\ref{eq:Z1}) and (\ref{eq:Z2})
reduce to
\begin{eqnarray}
(\Delta x)^2 &=& \frac{\gamma_0}{2}+\frac{\gamma_1}{2}-
\gamma^2 \;\cos^2 \Omega_1 t, \\
(\Delta p_x)^2 &=& \frac{\chi_0}{2}+\frac{\chi_1}{2}-
\eta^2 \;\sin^2 \Omega_1 t.
\end{eqnarray}

\begin{figure}
\begin{center}
\includegraphics[keepaspectratio=true,width=120mm]{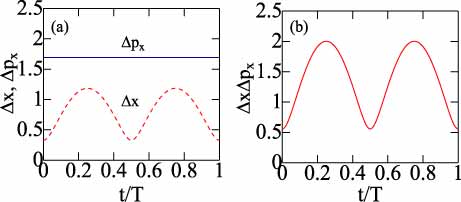}
\end{center}
\caption{
(Color online) 
(a) Time dependences of $\Delta x$ (dashed curve) 
and $\Delta p_x$ (solid curve), and
(b) the uncertainty: $\Delta x \Delta p_x$
for a linear coupling ($m=1.0$, $\alpha=1.0$, $c=1.0$ and $N=1$).
}
\label{fig18}
\end{figure}

Figure \ref{fig18}(a) shows time dependences of $\Delta x$ 
and $\Delta p_x$ for a linear coupling with $m=1.0$, $\alpha=10.0$, $c=1.0$ 
and $N=1$.
Although $\Delta x$ oscillates with an appreciable magnitude of $\gamma^2$ in Eq. (\ref{eq:Z1}),
$\Delta p_x$ is almost constant ($\simeq 1.69$) because of a small $\eta^2$ in Eq. (\ref{eq:Z2}).
Figure \ref{fig18}(b) shows the uncertainty given by 
$\Delta x \Delta p_x$, which is initially $0.556875$. 
The Heisenberg uncertainty relation: $\Delta x \Delta p_x \geq \hbar/2$ is always preserved.
We note that $\Delta x \Delta p_x$ has a large magnitude at $t \sim T/4$ or $3 T/4$
when tunneling takes place (Fig. \ref{fig5}).
This shows that the uncertainty is related with quantum tunneling \cite{Hasegawa13}.

\section{Concluding remark}
We have studied wavepacket dynamics in the Razavy hyperbolic DW potential 
\cite{Razavy80} which is coupled to a HO by linear and quadratic interactions.
Wavepackets show the quantum tunneling between two bottoms 
in the composite potential $U(x, y)$ [Eq. (\ref{eq:A15})].
The tunneling period is increased with increasing $c$ and/or $N$,
which is more significant for smaller $m$ and smaller $\alpha$
(Figs. \ref{fig15} and \ref{fig16}).
Comparing results of linear and quadratic couplings, we note that
the tunneling probability $P_r(t)$ is the same and
the marginal probability density $\rho_x(t)$ is similar between the two,
but $\rho_y(t)$ is different (Figs. \ref{fig6} and \ref{fig10}). 
Furthermore, $\langle y \rangle$ vs. $\langle p_y \rangle$ plot in the quadratic coupling 
stay at the initial position in the phase space,
while that in the linear coupling shows an elliptic motion (Figs. \ref{fig7} and \ref{fig11}).
Our phase space plots of $\langle x \rangle$ vs. $\langle p_x \rangle$
and  $\langle y \rangle$ vs. $\langle p_y \rangle$
for the two-term wavepacket given by Eq. (\ref{eq:A14})
are quite different from relevant results obtained in \cite{Babyuk02},
where trajectories show ellipse-like orbits but they
do not return to initial positions after revolution in a quartic DW potential coupled
to HO (see Figs. 4.2 and 4.5 in \cite{Babyuk02}).
Ref. \cite{Babyuk02} showed that this oddity occurs even for uncoupled DW
(see Fig. 3.2 in \cite{Babyuk02}), for which 
our calculation leads to the complete elliptic trajectory for $\langle x \rangle$ 
vs. $\langle p_x \rangle$ plot because we obtain
\begin{eqnarray}
\langle x \rangle &=& \gamma \:\cos \Omega_1 t,\;\;\;
\langle p_x \rangle= -\eta \:\sin \Omega_1 t,
\end{eqnarray}
for $c=0.0$ in Eqs. (\ref{eq:B9}) and (\ref{eq:B10}).
This difference between the result of Ref.\cite{Babyuk02} and ours does not arise
from the difference between quartic and hyperbolic DW potentials,
because a chain of equations of motion for expectation values in a general symmetric DW potential 
is closed within $\langle x \rangle$ and $\langle p_x \rangle$ for the two-term wavepacket
(see the Appendix). It has been shown that the uncertainty of $\Delta x \Delta p_x$ 
becomes appreciable when the tunneling takes place (Fig. 18).
It would be interesting to experimentally observed $\vert \Psi(x, t) \vert^2$ and
$\Delta x \Delta p_x$, which might be possible with advanced recent technology.

\begin{acknowledgments}
This work is partly supported by
a Grant-in-Aid for Scientific Research from 
Ministry of Education, Culture, Sports, Science and Technology of Japan.  
\end{acknowledgments}

\appendix*

\section{Expectation values for the two-term wavepacket}
\renewcommand{\theequation}{A\arabic{equation}}
\setcounter{equation}{0}

A coupled DW system is assumed to be described by the Hamiltonian given by 
\begin{eqnarray}
H=\frac{p_x^2}{2 M}+\frac{p_y^2}{2m}+U(x,y),
\end{eqnarray}
with
\begin{eqnarray}
U(x,y) &=& V(x)+\frac{m \omega^2 y^2}{2}-c \:x^d \:y,
\end{eqnarray}
where $V(x)$ denotes a general symmetric DW potential and $d=1$ ($d=2$)
signifies a linear (quadratic) coupling. 
We consider the two-term wavepacket given by 
\begin{eqnarray}
\Psi(x,y, t) &=& \frac{1}{\sqrt{2}}\large[\Phi_0(x, y) \:e^{-i E_0 t/\hbar}
+\Phi_1(x, y) \:e^{-i E_1 t/\hbar} \large],
\label{eq:Y0}
\end{eqnarray}
where real eigenfunctions of $\Phi_0(x, y)$ and $\Phi_1(x, y)$ satisfy 
the Sch\"{o}dinger equation 
\begin{eqnarray}
H \Phi_0(x,y) &=& E_0 \Phi_0(x,y), 
\label{eq:Y7}\\
H \Phi_1(x,y) &=& E_1 \Phi_1(x,y)
=(E_0+\Omega_1) \Phi_1(x,y).
\label{eq:Y8}
\end{eqnarray}
Expectation values of $\langle x \rangle$ and $\langle p_x \rangle$ for the wavepacket
are expressed by
\begin{eqnarray}
\langle x \rangle &=& a_x \:\cos \Omega_1 t, 
\label{eq:Y11}\\
\langle p_x \rangle &=& - b_x \:\sin \Omega_1 t,
\label{eq:Y12}
\end{eqnarray}
where
\begin{eqnarray}
a_x &=& \int \int \:\Phi_0(x,y) \:x \:\Phi_1(x,y)\:dxdy, \\
b_x 
&=& \int \int \:\Phi_0(x,y) \:(- i \partial_x) \Phi_1(x,y)\:dxdy.
\end{eqnarray}
Equations (\ref{eq:Y11}) and (\ref{eq:Y12}) lead to
\begin{eqnarray}
\frac{d \langle x \rangle}{d t} &=& - \Omega_1 a_x \:\sin \Omega_1 t 
= \left( \frac{\Omega_1 a_x}{b_x} \right) \langle p_x \rangle, 
\label{eq:Y1}\\
\frac{d \langle p_x \rangle}{d t} &=& - \Omega_1 b_x \:\cos \Omega_1 t
= - \left(\frac{\Omega_1 b_x}{a_x} \right) \langle x \rangle.
\label{eq:Y2}
\end{eqnarray}

On the other hand, Heisenberg equations of motion for $x$ and $p_x$ are given by
\begin{eqnarray}
\frac{d x}{d t} &=& \frac{\partial H}{\partial p_x}=\frac{p_x}{M}, 
\label{eq:Y3}\\
\frac{d p_x}{d t} &=& - \frac{\partial H}{\partial x}=-\partial_x U(x,y).
\label{eq:Y4}
\end{eqnarray}
Taking averages of Eqs. (\ref{eq:Y3}) and (\ref{eq:Y4}) over the two-term wavepacket 
$\Psi(x,y,t)$ given by Eq. (\ref{eq:Y0}), we obtain
\begin{eqnarray}
\frac{d \langle x \rangle}{d t} 
&=& - \frac{1}{M} \int  \int \Phi_0(x,y) \:\partial_x \Phi_1(x,y) \:dx dy\:\sin \Omega_1 t, 
\label{eq:Y5}\\
\frac{d \langle p_x \rangle}{d t} 
&=& - \int \int \Phi_0(x,y) \:\partial_x U(x,y) \:\Phi_1(x,y)\:dx dy \:\cos \Omega_1 t.
\label{eq:Y6}
\end{eqnarray}

The equivalence of Eqs. (\ref{eq:Y1}) and (\ref{eq:Y2}) with
Eqs. (\ref{eq:Y5}) and (\ref{eq:Y6}), respectively, may be shown as follows:
Multiplying Eqs. (\ref{eq:Y7}) and (\ref{eq:Y8}) by $x\:\Phi_1(x, y) $ 
and integrating them over $x$ and $y$ with integrations by parts, we obtain
\begin{eqnarray}
\Omega_1 a_x 
&=& \frac{1}{M} \int \int \Phi_0(x,y) \:\partial_x \Phi_1(x,y)\:dx dy.
\label{eq:Y9}
\end{eqnarray}
Multiplications of Eqs. (\ref{eq:Y7}) and (\ref{eq:Y8}) by $\partial_x \Phi_1(x, y)$ 
and integrations of them over $x$ and $y$ lead to
\begin{eqnarray}
\Omega_1 b_x 
&=& \int \int \Phi_0(x) \:\partial_x U(x,y) \:\Phi_1(x,y)\: dx dy.
\label{eq:Y10}
\end{eqnarray}
It is well known that Heisenberg equations of motion given by Eqs. (\ref{eq:Y3}) and (\ref{eq:Y4})
generally yield a hierarchical chain for DW potentials.
Fortunately, equations of motion for expectation values for the two-term
wavepacket close within $\langle x \rangle$ and $\langle p_x \rangle$ as given by
Eqs. (\ref{eq:Y1}) and (\ref{eq:Y2}). Such a simplification does not occur for
a general wavepacket, as given by Eqs. (\ref{eq:A13}) and (\ref{eq:A13b}).

Similar calculations may be made also for $\langle y \rangle$ and $\langle p_y \rangle$.
Then both the $\langle x \rangle$ vs. $\langle p_x \rangle$ plot
and the $\langle y \rangle$ vs. $\langle p_y \rangle$ plot are elliptic for the two-term
wavepacket.



\end{document}